\documentclass[a4paper,12pt]{article}
\usepackage{amsmath,amsfonts,amssymb,epsfig,subfigure,color}

\usepackage{fullpage}

\newcommand{\be}{\begin{equation}}
\newcommand{\en}{\end{equation}}

\renewcommand{\vec}[1]{\boldsymbol{#1}}

\begin{document}

\numberwithin{equation}{section}


\title{Simple shear is not so simple}


\author{Michel Destrade$^{ab}$, Jerry G. Murphy$^c$, Giuseppe Saccomandi$^d$,\\[12pt]
$^a$School of Mathematics, Statistics and Applied Mathematics,  \\ 
   National University of Ireland Galway, University Road, Galway, Ireland;\\[12pt]
 $^b$ School of Electrical, Electronic, and Mechanical Engineering,  \\
University College Dublin, Belfield, Dublin 4, Ireland;\\[12pt]
  $^c$Department of Mechanical Engineering,  \\ 
       Dublin City University, Glasnevin, Dublin 9, Ireland;\\[12pt]
$^d$Dipartimento di Ingegneria Industriale,  \\
Universit\`{a} degli Studi di Perugia,  06125 Perugia, Italy}

\date{}

\maketitle

\begin{abstract}
For homogeneous, isotropic, nonlinearly elastic materials, the form of the homogeneous deformation consistent with the application of a 
Cauchy shear stress is derived here for both compressible and incompressible materials. 
It is shown that this deformation is not simple shear, in contrast to the situation in linear elasticity. 
Instead, it consists of a triaxial stretch superposed on a classical simple shear deformation, for which the amount of shear cannot be greater than 1.
In other words, the faces of a cubic block cannot be slanted by an angle greater than $45^\circ$ by the application of a pure shear stress alone.
The results are illustrated for those materials for which the strain energy function does not depend on the principal second invariant of strain. 
For the case of a block deformed into a parallelepiped, the tractions on the inclined faces necessary to maintain the derived deformation are calculated.
\end{abstract}


\newpage


\section{Introduction}


Batra  \cite{Batr76} showed that in the framework of isotropic nonlinear elasticity a simple tensile load produces a {simple extension deformation if the empirical inequalities hold.  
Here we consider the related problem: if we apply a \emph{shear} stress to a nonlinearly elastic block, what deformation is produced? 
To formulate this question mathematically, we start by letting $\vec{x}(\vec{X})$ denote the current position of a particle which was located at $\vec{X}$ in the reference configuration. 
Denote the unit vectors associated with a fixed Cartesian coordinate system  in the reference configuration by $(\vec{E}_1,\vec{E}_2,\vec{E}_3)$ and in the current configuration by $(\vec{e}_1,\vec{e}_2,\vec{e}_3)$. Consider now a rectangular block of a non-linearly elastic material, with edges aligned with  $\vec{E}_1$, $\vec{E}_2$, and $\vec{E}_3$. 
Batra \cite{Batr76} was interested in obtaining  deformations  $\vec{x}(\vec{X})$ associated with a Cauchy stress tensor of the form
\be \label{tense}
\vec{\sigma}= T(\vec{e}_1\otimes\vec{e}_{1}), 
\en
where $T$ is a constant,
whereas here the concern is determining the 
deformations associated with a Cauchy stress of the form
\begin{equation} \label{as}
\vec{\sigma}
 = S(\vec{e}_1\otimes\vec{e}_{2}+\vec{e}_2\otimes\vec{e}_{1}),
\end{equation}
where $S$ is a constant (in order to ensure that the equations of equilibrium are satisfied).
We call \eqref{as} a uniform \emph{shear stress} \cite{Spencer, Holz}, although it is also denoted  ``pure shear''   in the literature, e.g. \cite{Lurie, Drozdov, Atkin, Gurtin, Gonz08}.

The problem being considered is therefore the \emph{stress} formulation of the problem of shear (the ``most illuminating homogeneous static deformation" according to Truesdell \cite{True}). 
The approach adopted here is the inverse of the classical \emph{strain} formulation of the same problem, first proposed by Rivlin \cite{Rivlin}. In that approach, the relative parallel motion of two opposite faces of a block is considered. 
Motivated by linear elasticity and his own physical insight, Rivlin proposed that the resulting deformation of the block can be described by
\begin{equation} \label{ssstrain}
x=X+ KY, \qquad y=Y, \quad z=Z,
\end{equation}
where $(X,Y,Z)$ and $(x,y,z)$ are the Cartesian coordinates of a typical particle before and after deformation, respectively, and  $K$ is the amount of shear. 
Using his now classical constitutive theory, Rivlin was able to calculate the stress distribution necessary to support this deformation. 
This strain formulation of the simple shear problem has since received much attention in the literature with excellent summaries to be found, for example, in Atkin and Fox \cite{Atkin} and Ogden \cite{Ogden}. 

Note that in the \emph{linear} theory, these two formulations are equivalent since an infinitesimal shear stress of the form \eqref{as} leads to an infinitesimal deformation of the form \eqref{ssstrain} and vice versa. 
As noted by Rivlin \cite{Rivlin},  this equivalence does not carry over to the framework of \emph{finite} elasticity. 
We address this problem here. 

We recall that the dead-load traction boundary condition in finite elasticity is the prescription of the first Piola-Kirchhoff stress vector on the boundary of the undeformed body.
Here we focus on the application of a \emph{Cauchy} shear stress, because it gives a one-to-one mapping between a $<$deformation, traction$>$ pair and a stress vector on a boundary. By contrast, the assignment of a Piola-Kirchhoff traction leads to a many-to-one mapping and has thus an ambiguous physical meaning \cite{Hoge91}.


\section{Invertibility of the stress-strain relation}


Let $\vec{B}$ denote the left Cauchy-Green strain tensor. A homogeneous, isotropic, hyperelastic  material possesses a strain-energy density which may be written as $W = W(I_{1},I_{2}, I_{3})$, where $I_{1},I_{2},I_{3}$ are the three principal invariants of $\vec{B}$, given by
\be
\label{T-B}
I_{1}=\text{tr} \ \vec{B}, \qquad I_{2}= \textstyle{\frac{1}{2}} \left[I_1^2-\text{tr}(\vec{B}^2) \right],
\qquad
I_{3}=\det \vec{B}.
\en
The general representation formula for the Cauchy stress tensor $\vec{\sigma}$ is then
\begin{equation} \label{repres}
\vec{\sigma}=\beta_0 \vec{I}+\beta_1 \vec{B} + \beta_{-1} \vec{B}^{-1},
\end{equation}
where   
\be \label{ssrel}
\beta_0=\frac{2}{I_3^{1/2}} \left[I_2\dfrac{\partial W}{\partial I_2}+I_3 \dfrac{\partial W}{\partial I_3} \right], \qquad \beta_1= \frac{2}{I_3^{1/2}}\dfrac{\partial W}{\partial I_1}, \qquad 
\beta_{-1}=-2I_3^{1/2} \dfrac{\partial W}{\partial I_2}.
\en
It is assumed that the so-called empirical inequalities hold,
\be \label{emin}
\beta_0 \leq 0, \qquad \beta_1 > 0, \qquad \beta_{-1} \leq 0.
\en

If $W$ is independent of $I_2$, then  $\beta_{-1}=0$ and it follows from \eqref{repres} that
\be \label{inv1}
\vec{B} = -\dfrac{\beta_0}{\beta_1}\vec{I} + \dfrac{1}{\beta_1}\vec{\sigma}.
\en
Assume now that $\beta_{-1}<0$. Then Johnson and Hoger \cite{Hoger} have shown that 
\begin{equation} \label{inv2}
\vec{B}=\psi_0 \vec{I}+\psi_1 \vec{\sigma} + \psi_2 \vec{\sigma}^2,
\end{equation}
where
\begin{align}
& \psi_0 = \dfrac{1}{\Delta}\left(\beta_0^2 - 2\beta_{-1}\beta_1 + I_2\beta_1^2 + I_3\dfrac{\beta_0\beta_1^2}{\beta_{-1}} + I_1 \beta_{-1}^2 + \dfrac{I_2}{I_3}\beta_0 \beta_{-1}\right), \notag \\
& \psi_1 = -\dfrac{1}{\Delta}\left(2\beta_0 + I_3\dfrac{\beta_1^2}{\beta_{-1}} +\dfrac{I_2}{I_3}\beta_{-1}\right), \notag \\
& \psi_2 = \dfrac{1}{\Delta},\notag \\
& \Delta = I_1\beta_1^2 - I_3\dfrac{\beta_1^3}{\beta_{-1}} + \dfrac{1}{I_3} \beta_{-1}^2 - \dfrac{I_2}{I_3}\beta_1 \beta_{-1}.
\end{align}
These coefficients are all positive by virtue of the empirical inequalities \eqref{emin},
\be
\psi_0>0, \qquad \psi_1>0, \qquad \psi_2>0.
\label{psi-sign}
\en 

For incompressible materials, only isochoric deformations are admissible and $I_3=1$ at all times. Then $W=W(I_1,I_2)$ only, and the representation formula is given by     
\begin{equation}
\vec{\sigma}=-p\vec{I}+\beta_1 \vec{B}+\beta_{-1}  \vec{B}^{-1},
\label{2}
\end{equation}
where $p$ is the indeterminate Lagrange multiplier introduced by the constraint of incompressibility and 
\be \label{inco}
\beta_1 = 2 \dfrac{\partial  W}{\partial I_1}, 
\qquad 
\beta_{-1} = - 2 \dfrac{\partial W}{\partial I_2}.
\en
The empirical inequalities in this case have the form 
\be \label{eminin}
 \beta_1 > 0, \qquad
 \beta_{-1} \leq 0.
\en
The inverse form of this relation is given by \eqref{inv1}, \eqref{inv2} with $\beta_0$ replaced by $-p$.

This invertibility of the classical stress-strain relation for both compressible and incompressible materials plays a central role in what follows.


\section{Determination of the strain}


The proof that a uni-axial tension load produces an equi-biaxial contraction in isotropic finite elasticity is due to Batra \cite{Batr76}. An alternative proof is given here, based on the invertibility relations of the last section. First note that a stress distribution of the form \eqref{tense}, i.e.
\be
\vec{\sigma} = \begin{bmatrix} T & 0 & 0 \\ 0 & 0 & 0 \\ 0 & 0 & 0 \end{bmatrix},
\en
leads to the equi-axial deformation
\be \label{bcom}
\vec{B} = 
\begin{bmatrix}
B_{11} & 0 & 0
\\
0 & B_{33} & 0
 \\
0 & 0 & B_{33}
 \end{bmatrix},
\en
as expected, where, if $\beta_{-1}<0$,
\be \label{stcom1}
B_{11} = \psi_0 + \psi_1T + \psi_2T^2, \qquad B_{33}= \psi_0, 
\en
and, if $\beta_{-1}=0$,
\be \label{stcom2}
 B_{11}=\left(T-\beta_0\right)/\beta_1, \qquad B_{33}=-\beta_0/\beta_1.
\en
In both cases, $B_{11} - B_{33}$ is of the same sign as $T$, owing to \eqref{psi-sign} and \eqref{eminin}, which shows that \emph{universally}, a uni-axial tension (compression) leads to a lateral contraction (expansion).

Following the same line of reasoning, we see that a Cauchy shear stress distribution of the form \eqref{as}, i.e.
\be
\vec{\sigma} = \begin{bmatrix} 0 & S & 0 \\ S & 0 & 0 \\ 0 & 0 & 0 \end{bmatrix},
\en
leads to a left Cauchy-Green strain tensor of the form
\be \label{B-general}
\vec{B} = 
\begin{bmatrix}
B_{11} & B_{12} & 0
\\
B_{12} & B_{11} & 0
 \\
0 & 0 & B_{33}
 \end{bmatrix},
\en
where 
\be \label{coeff}
B_{11}= \psi_0 + \psi_2 S^2, 
\qquad 
B_{12} = \psi_1 S, 
\qquad
B_{33} = \psi_0.
\en
if $\beta_{-1}<0$ and, if $\beta_{-1}=0$,
\be \label{spec}
B_{11}=B_{33}=-\beta_0/\beta_1, \qquad B_{12}=S/\beta_1.
\en

This is \emph{not} compatible with the left Cauchy-Green strain tensor associated with the simple shear deformation \eqref{ssstrain} which has the form
\be \label{B-simple-shear}
\vec{B} = 
\begin{bmatrix}
1+K^2 & K & 0
\\
K & 1 & 0
 \\
0 & 0 & 1
 \end{bmatrix}.
\en

The strain \eqref{coeff} combines triaxial stretch with simple shear, a type of deformation which has been investigated previously by Payne and Scott \cite{Payne}, Varga \cite{Varga}, Wineman and Gandhi \cite{WiGa}, Rajagopal and Wineman \cite{RaWi}, and Destrade and Ogden \cite{DeOg05}. 
To make this connection clear, we first note that $B_{11}>0$, $B_{11}^2-B_{12}^2>0$, $B_{33}>0$, because the strain tensor is positive definite. 
It follows that \eqref{B-general} can be written as 
\be 
\vec{B} = 
\begin{bmatrix}
\lambda_2^2 & \lambda_2^2\sqrt{1 - \lambda_1^2\lambda_2^{-2}} & 0
\\[4pt]
 \lambda_2^2\sqrt{1 - \lambda_1^2\lambda_2^{-2}} & \lambda_2^2 & 0
 \\
0 & 0 & \lambda_3^2
 \end{bmatrix},
\en
where $\lambda_1 = \sqrt{(B_{11}^2-B_{12}^2)/B_{11}}$, $\lambda_2 = \sqrt{B_{11}}$, and $\lambda_3=\sqrt{B_{33}}$.
(Here, and henceforth, we assumed that $S>0$. In the case $S<0$, a minus sign must be inserted in front of the square root.)
A simple check shows that $\vec{B} = \vec{F F}^T$ where the deformation gradient $\vec{F}$ can be decomposed as 
\be \label{F}
\vec{F} = 
\begin{bmatrix}
1 & \sqrt{1 - \lambda_1^2\lambda_2^{-2}} & 0
\\[4pt]
 0 & 1 & 0
 \\
0 & 0 & 1
 \end{bmatrix}
 \begin{bmatrix}
\lambda_1 & 0 & 0
\\
 0 & \lambda_2 & 0
 \\
0 & 0 & \lambda_3
 \end{bmatrix},
\en
corresponding to a triaxial extension with principal stretch ratios $\lambda_1$, $\lambda_2>\lambda_1$, $\lambda_3$, followed by a simple shear with amount of shear $\sqrt{1 - \lambda_1^2\lambda_2^{-2}}$. 
Here we notice a surprising \emph{universal result}: any isotropic, hyperelastic material cannot be shared by an amount of shear greater than 1, independently of the magnitude of the shear stress;
this limit corresponds to a maximum shear angle of $\tan^{-1}(1) = 45^\circ$. 

The corresponding deformation is given by
\be \label{def}
x= \lambda_1X+\lambda_2\sqrt{1 - \lambda_1^2\lambda_2^{-2}} \,Y, 
\qquad
y=\lambda_2Y, 
\qquad
z=\lambda_3Z,
\en
for compressible materials. The corresponding incompressible form is given by
\be \label{defin}
x= \lambda_1X+\lambda_2\sqrt{1 - \lambda_1^2\lambda_2^{-2}} \, Y, 
\qquad
y=\lambda_2Y, 
\qquad
z=\lambda_1^{-1}\lambda_2^{-1}Z.
\en

The principal stretches of the deformation \eqref{def}, $\mu_1, \mu_2$ and $\mu_3$ (say) can be computed from the general expressions found in \cite{RaWi, DeOg05}.
Here they are connected to the $\lambda$'s through
\be
\lambda_1^2 = \dfrac{\mu_1^2+\mu_2^2}{2 \mu_1 \mu_2}, \qquad \lambda_2^2=\mu_1 \mu_2,
\qquad \mu_3=\lambda_3.
\en
For infinitesimal strains, $\mu_i=1 + \epsilon_i$, say, and the amount of shear $K=\sqrt{1-\lambda_1^2\lambda_2^2}$ becomes $K \simeq (\epsilon_1+\epsilon_2)^{1/2}$. 
At the lowest order in $K$ (order 1), we find that $\lambda_1 \simeq 1$, $\lambda_2  \simeq 1$, $\lambda_2 K \simeq K$, $\lambda_3 \simeq 1$ 
and thus, that the deformations \eqref{def} and \eqref{defin} coincide with the small strain simple shear deformation \eqref{ssstrain}.

Finally, we note that since we have an expression for the deformation gradient $\vec{F}$ in \eqref{F}, it is a simple matter to compute the components of the first Piola-Kirchhoff stress tensor $\vec P = \left(\det \vec F \right) \vec T \left(\vec{F}^{-1}\right)^T$ corresponding to the shear stress \eqref{as}.
We find that 
\be 
\vec{P} = 
\begin{bmatrix}
0 & \lambda_1\lambda_3S & 0 \\
\lambda_2 \lambda_3 S & -\lambda_2\lambda_3 \sqrt{1 - \lambda_1^2\lambda_2^{-2}} \, S & 0
\\ 0 & 0 & 0
 \end{bmatrix}.
 \en


\section{Special materials}


A wide class of isotropic materials is modeled by a strain-energy density $W$ which does not depend on $I_2$, the second principal invariant of strain. 
This leads to $\beta_{-1}=0$ and the stress-strain inversion is given by \eqref{inv1}.

It follows that $B_{33} = B_{11}$ in \eqref{B-general} and, further, that $\lambda_2=\lambda_3$ in \eqref{F}. In other words, for materials such that $\partial W / \partial I_2 = 0$, a shear stress \eqref{as} produces a simple tension in the shear direction (with tensile extension 
$\lambda_1$ and lateral contraction $\lambda_2$), followed by a simple shear of amount $\sqrt{1 - \lambda_1^2\lambda_2^{-2}}$, yielding a deformation of the form
\be \label{defsp}
x= \lambda_1X+\lambda_2\sqrt{1 - \lambda_1^2\lambda_2^{-2}}\, Y, 
\qquad
y=\lambda_2Y, 
\qquad
z=\lambda_2Z.
\en
This is a two-parameter deformation, entirely determined by $\lambda_1$ and  $\lambda_2$. These quantities are related to the shear stress and the material response functions as follows,
\be
\lambda_1 = \sqrt{\dfrac{S^2 - \beta_0^2}{\beta_0\beta_1}}, \qquad
\lambda_2 = \sqrt{-\dfrac{\beta_0}{\beta_1}},
\en
where $\lambda_2$ is real by the empirical inequalities.

In the incompressible case (the so-called \emph{generalized neo-Hookean materials}), $W = W(I_1)$ only, where $I_1$ is the first principal invariant. 
Then we have a {one}-parameter deformation,
\be \label{spin}
x=\lambda_1X+\sqrt{\frac{1-\lambda_1^3}{\lambda_1}}\, Y, 
\qquad
y = \lambda_1^{-1/2}Y, 
\qquad
z = \lambda_1^{-1/2}Z,
\en
with
\be \label{F-neo}
\vec{F} = 
\begin{bmatrix}
1 & \sqrt{1 - \lambda_1^3} & 0
\\[4pt]
 0 & 1 & 0
 \\
0 & 0 & 1
 \end{bmatrix}
 \begin{bmatrix}
\lambda_1 & 0 & 0
\\
 0 & \lambda_1^{-1/2} & 0
 \\
0 & 0 & \lambda_1^{-1/2}
 \end{bmatrix},
\en
and
\be
\lambda_1 = \sqrt{\dfrac{p^2-S^2}{2pW'}}.
\en
Here, $p$ is the Lagrange multiplier introduced by the constraint of incompressibility.
Note that the strain-stress relation is now
\be
\vec{B} = \dfrac{p}{2W'}\vec{I} + \dfrac{1}{2W'}\vec{\sigma}.
\en
This yields, in particular, that $p=2W'B_{33}$, which is positive by the empirical inequalities and the positive definitiveness of $\vec{B}$. 

Notice also from \eqref{spin} that, in order for the deformation to be well-defined, $0<\lambda_1<1$.
Therefore for \emph{all} generalized neo-Hookean materials, a shear stress leads to a simple shear combined with a uni-axial \emph{compression} in the direction of shear (and thus, a bi-axial \emph{expansion} in the transverse plane). 
In other words, the gliding plane $Y = $const. is ``lifted'' by the application of the shear stress \eqref{as},  as can be seen in Fig. 1 (and the plane of shear $Z = $ const. is pushed  ``outwards''. 

\begin{figure}[ht]
\begin{center}
\includegraphics[width=0.56\textwidth]{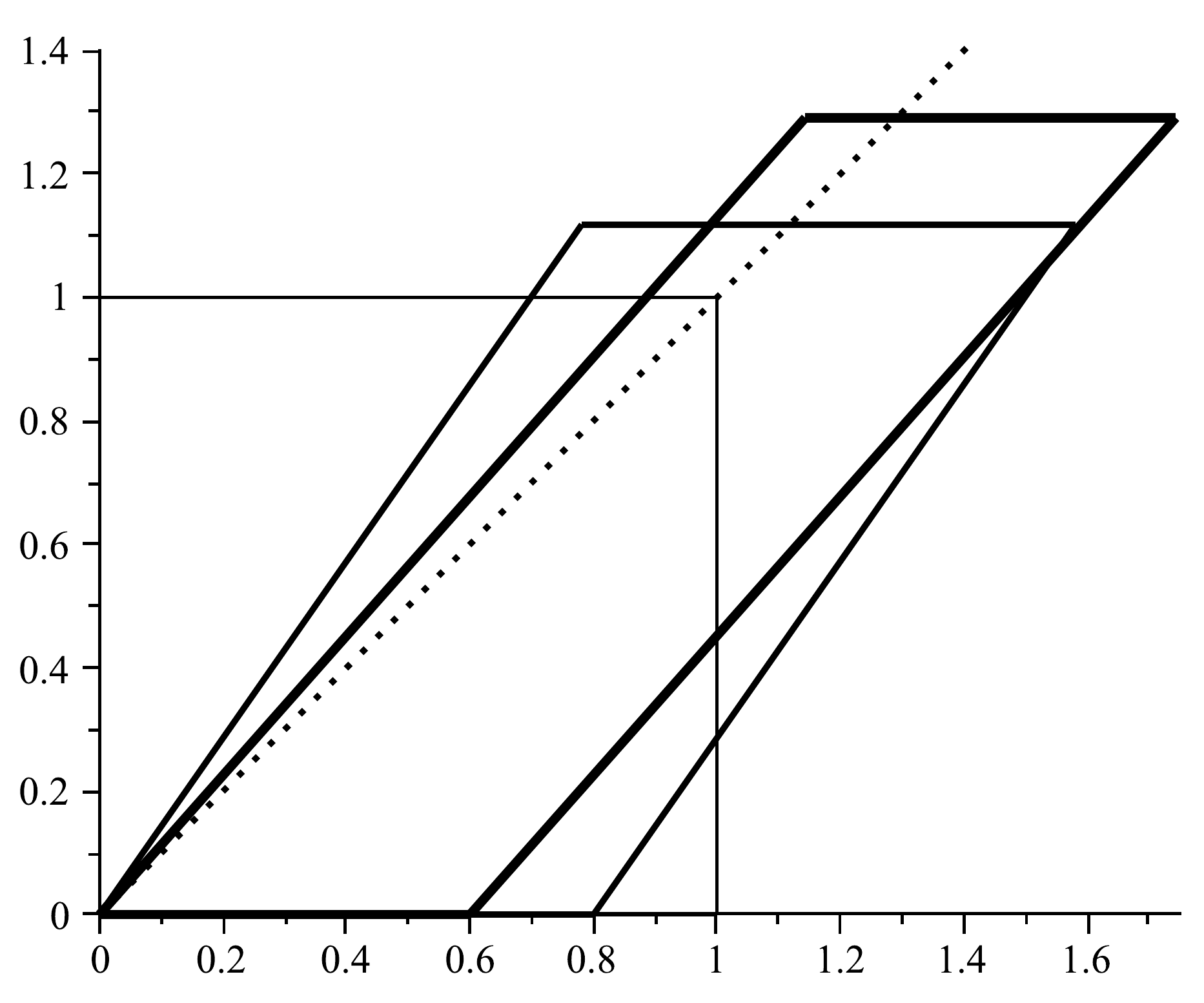}
\caption{Cross-section of a unit block made of any generalized neo-Hookean material, subjected to an increasing shear stress of the form $S(\vec{e_1}\otimes\vec{e_2} + \vec{e_2}\otimes\vec{e_1})$, where the $\vec{e_i}$ are aligned with the coordinate axes. 
The resulting deformation is a simple shear combined with a uni-axial compression in the direction of shear (here, successively of 0\%, 20\%, and 40\% along $\vec{e_1}$).
The face initially at $X=0$ cannot be inclined beyond the diagonal dotted line at $45^\circ$.}
\end{center}
\end{figure}


\section{Normal tractions}


Somewhat counterintuitively in view of the assumed stress distribution \eqref{as}, if one deforms a cuboid (with faces at $X = \pm A$, $Y = \pm B$, $Z = \pm C$, say)  into a parallelepiped, as is usual when visualising simple shear, \emph{normal} as well as shear tractions have to be applied to the inclined faces in order to maintain the homogenous deformation \eqref{def}. 

It follows easily from \eqref{def} that the unit normal, $\vec{n}$, in the current configuration to the face originally at $X=A$ is given by
\be \label{nor}
\vec{n}=\left(n_1,n_2,0\right)=\frac{1}{\sqrt{2 - \lambda_1^2\lambda_2^{-2}}}\left(1,-\sqrt{1 - \lambda_1^2\lambda_2^{-2}},0\right).
\en
Let $\vec{s}$ denote the tangent vector  defined by
\be \label{tan}
\vec{s}=\left(-n_2,n_1,0\right).
\en
The normal and shear tractions on the inclined face, originally at $X=A$ in the undeformed configuration, that are therefore necessary to maintain the deformation \eqref{def} are given respectively by
\begin{align} \label{tra}
& n =\vec{n \cdot \sigma}\vec{n}=2 S n_1 n_2
 = -2 S \frac{\sqrt{1 - \lambda_1^2\lambda_2^{-2}}}{2 - \lambda_1^2\lambda_2^{-2}}, 
\nonumber \\
& s =\vec{s \cdot \sigma}\vec{n}=S\left(n_1^2-n_2^2\right) = S\frac{\lambda_1^2\lambda_2^{-2}}{2 - \lambda_1^2\lambda_2^{-2}}.
\end{align}
It follows immediately from these that 
\be
-2S < n \leq 0, 
\qquad
0 \leq s < S,
\en
and so the necessary normal traction on the slanted face is always compressive.

For  generalised neo-Hookean materials, $\lambda_2=\lambda_1^{-1/2}$, and these tractions therefore simplify to the following:
\be \label{tranh}
\hat{n} \equiv n/S = -2\frac{\sqrt{1-\lambda_1^3}}{2 - \lambda_1^3}, 
\qquad
\hat{s} \equiv s/S = \frac{\lambda_1^3}{2-\lambda_1^3}.
\en
These forms are independent of the specific form of the strain-energy function and are plotted in Figure 2. Note, in particular, the rapid decay of the normal traction once loading begins. This suggests that a normal compressive traction of the same order as the shear stress has to applied to maintain homogeneity of the deformation once shear loading begins. Application of the shear traction is not as important.
\begin{figure}
\begin{center}
\includegraphics[width=0.8\textwidth]{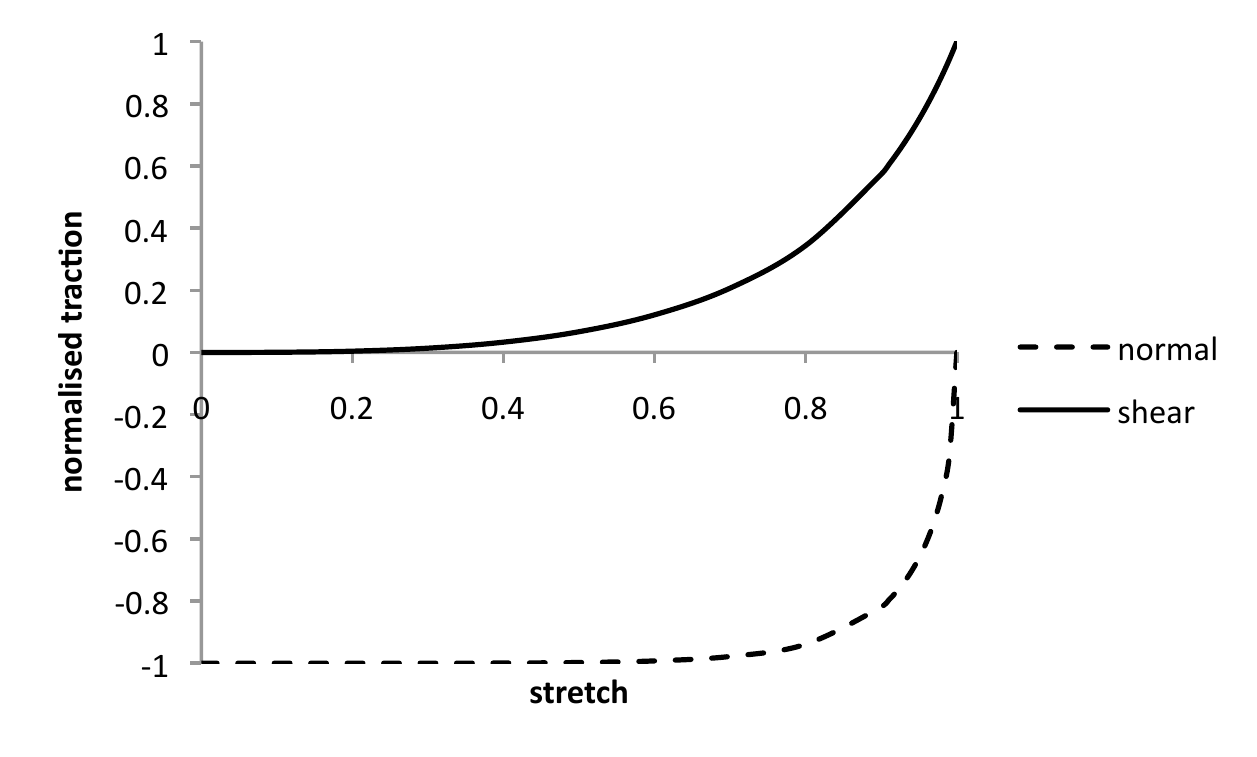}
\caption{Plots of the normal and shear components of the traction on slanted face versus $\lambda_1$.}
\end{center}
\end{figure}

There are two important geometrical idealisations where these tractions on the inclined faces can be ignored. 
The first idealisation is that of a thin block, for which $B \gg A$. 
Here applying a constant shear stress to the faces $Y = \pm B$ generates the deformation \eqref{def}. The second important idealisation is that of a half-space, with the semi-infinite dimension in the $Y$ direction. 
Application of a constant shear stress then at $Y=0$ (and the generation of an equilibriating shear stress at infinity) again yields \eqref{def}.

Finally we conclude by recalling that Lurie \cite{Lurie} shows that for the shear stress \eqref{as}, the two in-plane principal stresses are aligned with the diagonals $X \pm Y = $ const. and have both magnitude $S/2$.



\section{Concluding Remarks}


Although all textbooks on nonlinear elasticity explain that a simple shear cannot be produced by shear forces alone, very few allude to what deformation is produced by a pure shear stress. 
In fact, we found only three such mentions, two correct: ``In an isotropic non-linear elastic medium the state of pure shear is not accompanied by pure shear strain'' \cite{Lurie} and ``Pure shear is a significantly more complicated deformation compared with simple shear'' \cite{Drozdov} and one incorrect: ``A state of uniform shear stress leads to a uniform (or simple) shear deformation'' \cite{Holz}.
In this note we derived the homogeneous deformation consistent with the application of a shear stress. 

We wish to emphasize that this is not a purely academic problem. Indeed, in a laboratory it is the applied force that is easily controlled, not the deformation. 
The displacement in simple shear may be controlled only in special devices and in special situations. 
As Beatty \cite{Beatty} put it: ``In practice, [..] it is not likely that a global simple shear deformation may be produced in any real material''.
Similarly, Brown explains in his handbook \emph{Physical Testing of Rubber} \cite{Brown} that ``With single and double sandwich construction, there is a tendency for the supporting plates to move out of parallel under load'', a phenomenon which clearly corrupts the concept of simple shear. 
As a remedy, the British Standard ISO 1827:2007 \cite{ISO} advocates the use of quadruple shear test devices. 
In those tests, however, the distance between the plates is allowed to vary freely, and again, the simple shear deformation is modified.

Therefore, the determination of the deformation field corresponding to a  simple shearing stress field, in the framework of isotropic nonlinear elasticity, should be useful in the designing experimental protocols for real materials. 
We have shown that a pure shearing stress field produces a deformation field which is \emph{not} the classical simple shear deformation, but a simple shear deformation superimposed upon a triaxial stretch, and that the corresponding angle of shear cannot be greater than $45^\circ$. 
For the special case of incompressible generalized neo-Hookean materials,  a one-parameter isochoric deformation is obtained. 
For general isotropic materials the corresponding deformation field is more complex and even more complex for anisotropic materials (see, for example, \cite{BeattySacco, DGPS08}).

\section*{Acknowledgements}
The motivation for this paper originates from discussions at a conference  held in honour of K.R. Rajagopal in November 2010, and we are most grateful to Texas A\&M University for its generous support. MD also acknowledge support from the NUI Galway Millenium Fund to attend that conference.

We thank Alain Goriely, Michael Hayes, Cornelius Horgan, and the referees for their positive comments on an earlier version of the paper. 
In particular, we are most grateful to Alain Goriely for bringing to our attention a little-known (or at least, little-cited) paper by Moon and Truesdell \cite{MoTr74}, which has a large overlap with this one. 
Specifically, the result about a maximum amount of shear in shear stress can be found there, and we cannot thus claim novelty here;
our paper confirms and complements the work of Moon and Truesdell, especially in the light of subsequent developments such as the triaxial strech and shear decomposition \cite{WiGa, RaWi}.


\end{document}